\documentclass[10pt]{article}
\usepackage{amsmath,amsfonts,xypic}
\usepackage{amssymb}
\usepackage{amscd}
\usepackage[all]{xy}
\advance \textheight by \topskip
\makeatletter

\newtheorem{theorem}{Theorem}[section]

\newtheorem{definition}{Definition}[section]

\newcommand{\beqa}{\begin{eqnarray}}
\newcommand{\eeqa}{\end{eqnarray}}
\newcommand{\noi}{\noindent}

\newcommand{\tom}{\tilde \Omega}

\newcommand{\cI}{{\cal{I}}}
\newcommand{\ccR}{{\cal{R}}}
\newcommand{\cG}{{\cal{G}}}

\begin{document} 
 
\title{Feynman amplitudes in renormalizable non-commutative quantum field theory} 
\author{ 
 A. Tanasa${}^{(1), (2)}$\footnote{e-mail: 
  adrian.tanasa@ens-lyon.org}\\ 
${}^{(1)}$Laboratoire de Physique Th\'eorique, B\^at.\ 210, CNRS UMR 8627\\ 
    Universit\'e Paris XI,  F-91405 Orsay Cedex, France\\ 
${}^{(2)}$ Dep. Fizica Teoretica, Institutul de Fizica si Inginerie Nucleara 
    H. Hulubei,\\ 
    P. O. Box MG-6, 077125 Bucuresti-Magurele, Romania 
} 
\maketitle 
 
\begin{abstract} 
We consider here the Feynman amplitudes of renormalizable non-commutative
quantum field theory models. Different representations (the parametric and the
Mellin one) are presented. The latter further allows the proof of meromorphy
of a amplitude in the space-time dimension.
\end{abstract}

\section{Introduction}
\setcounter{equation}{0}
 
This paper presents different representations of Feynman amplitudes of 
non-commutative quantum field theory (NCQFT) models. The models considered here are
the renormalizable models, namely the Grosse-Wulkenhaar model \cite{GW1, GW2}
or the ``covariant models'' (which include the non-commutative Gross-Neveu
model or the Langmann-Szabo-Zarembo - LSZ - model).

%UV-IR mixing, divergence which is responsible for the non-renormalizability of 
%the models. This difficulty was overtaken for scalar $\Phi^4$ models by the 
%introduction of a new harmonic term in the action - the Grosse-Wulkenhaar 
%model. 
% (see subsection \ref{GW}).  
%The model was proven to be 
%renormalizable at any order in perturbation theory \cite{GW1, GW2, GW5,  
%  GW3, GW4}. Moreover, the parametric representation was introduced \cite{param1} 
%and then the dimensional regularization and renormalization were performed  
%\cite{dimreg}. Let us also emphasize here that the Hopf algebra 
%description of this type of renormalization was given in 
%\cite{hopf}. Moreover, let us also stress here on the fact that it was recently shown 
%\cite{ultimul} that this type of action can be interpreted from the spectral 
%action (for latest developments see \cite{spectral}) point of view. 

\section{Renormalizable NCQFT models}
\setcounter{equation}{0}
  
We consider the $4-$dimensional Moyal space
%\beqa
%\label{2D}
 $[x^\mu, x^\nu]=i \Theta^{\mu \nu}, $ 
%\eeqa
%\noi
where the the matrix $\Theta$ is
\begin{eqnarray}
\label{theta}
  \Theta= 
  \begin{pmatrix}
 %   \begin{matrix} 
0 &\theta & 0 & 0\\ 
%      \hspace{-.5em} -\theta & 0
%    \end{matrix}    &&     0
%    \\ 
%    &\ddots&\\
%    0&&
%    \begin{matrix}0&\theta\\
%      \hspace{-.5em}-\theta & 0
%    \end{matrix}
-\theta & 0 & 0 & 0\\
0 & 0 & 0 & \theta \\
0 & 0 & -\theta & 0
  \end{pmatrix}.
\end{eqnarray}
\noi
%The associative Moyal product of two functions 
%$f$ and $g$ 
%on the Moyal space writes
%\beqa
%\label{moyal-product} 
% (f\star g)(x)=\int \frac{d^{4}k}{(2\pi)^{4}}d^{4}y\, f(x+{\textstyle\frac 12}\Theta\cdot
%  k)g(x+y)e^{\imath k\cdot y}\nonumber\\
%  &=&\frac{1}{\pi^{D}|\det\Theta|}\int d^{D}yd^{D}z\,f(x+y)
%  g(x+z)e^{-2\imath y\Theta^{-1}z}\; .
%\eeqa
We also consider an Euclidean metric. Let us now introduce the two types
of renormalizable non-commutative models.

\subsection{The Grosse-Wulkenhaar model}

%The action introduced in \cite{GW2} is 
%\beqa
%\label{lag-init}
%S=\int d^4 x \left(\frac{1}{2} \partial_\mu \phi
%\star \partial^\mu \phi +\frac{\Omega^2}{2} (\tilde{x}_\mu \phi )\star
%(\tilde{x}^\mu \phi ) 
%+ \frac{1}{2} m^2 \,\phi \star \phi
%+  \phi \star \phi \star \phi \star\phi \right)
%\eeqa
%\noi

%The propagator of this model is the inverse of the operator
%\beqa
%\label{inverse-propa}
%\Delta+\Omega^{2}\tilde x^{2}.
%\eeqa
%\noi
The results  established in the sequel hold for models
with interactions of type $\bar \phi \star \phi \star \bar \phi \star
\phi$. One has the Grosse-Wulkenhaar
model of a complex scalar field
\beqa
\label{lag}
S_{GW}=\int d^4 x \left( \partial_\mu \bar \phi
\star \partial^\mu \phi +{\Omega^2} (\tilde{x}_\mu \bar \phi )\star
(\tilde{x}^\mu \phi ) 
+  \bar \phi \star \phi \star \bar \phi \star \phi \right)
\eeqa
where 
%\beqa
%\label{tildex}
 $\tilde{x}_\mu = 2 (\Theta^{-1})_{\mu \nu} x^\nu.$ 
%\eeqa
%\noi
%One obtains for the propagator of a
This action leads to the following  propagator from a point $x$ to a point $y$:
\begin{equation}
\label{propa1}
C(x,y)=\int_0^\infty \frac{\tilde \Omega d\alpha}{[2\pi\sinh(\alpha)]^{2}}
e^{-\frac{\tilde \Omega}{4}\coth(\frac{\alpha}{2})(x-y)^2-
\frac{\tilde \Omega}{4}\tanh(\frac{\alpha}{2})(x+y)^2}\; .
\end{equation} 
Let us now introduce the {\it short} and {\it long variables}:
%\beqa
%\label{def-uv}
$u=\frac{1}{\sqrt 2} (x-y)$  and $ v=\frac{1}{\sqrt 2} (x+y)$.
%\eeqa
 Let 
%\beqa
%\label{t}
 $t_\ell = {\rm tanh} \frac{\alpha}{2}.$
%\eeqa
The propagator \eqref{propa1} becomes
\begin{equation}
\label{propa}
C(x,y)=\int_0^\infty \frac{\tilde \Omega d\alpha}{[2\pi\sinh(\alpha)]^{2}}
e^{-\frac{\tilde \Omega}{2}\frac{1}{t_\ell} u^2-
\frac{\tilde \Omega}{2}t_\ell v^2}\; .
\end{equation}

\subsection{The covariant models}

As already stated in the introduction, amongst this type of models one has the
non-commutative Gross-Neveu model and the LSZ model. The
results we present in the sequel hold for the latter but they can be
however extended for the Gross-Neveu model also. The
LSZ action writes:
\beqa
\label{lag2}
S_{LSZ}=\int d^4 x \left( (\partial_\mu \bar \phi - i \Omega \tilde x_\mu\phi)
\star (\partial^\mu  \phi -i \Omega  \tilde x^\mu \phi)
 + \bar \phi\star\phi\star\bar \phi\star\phi \right).
\eeqa
This action leads to the propagator
\beqa
\label{propa2}
C(x,y)=2 \int_0^1 d t_\ell \frac{\tilde \Omega(1-t_\ell^2)}{(4\pi
  t_\ell)^2}e^{-\frac 12 \tom \frac {1+ t_\ell^2}{2 t_\ell}u^2 + i \tom  u\wedge v},
\eeqa
where
$ u\wedge v = u_1 v_2 - u_2 v_1 + u_3 v_4 - u_4 v_3.$

\subsection{Non-local interaction}

Using the explicit form 
%\eqref{moyal-product} 
of the Moyal product, the interaction
term of both \eqref{lag} and \eqref{lag2} 
lead to the following contribution in position space
\beqa
\label{v1}
%V(x_1^V, x_2^V, x_3^V, x_4^V)=
\delta (x_1^V - x_2^V + x_3^V - x_4^V)e^{2i\sum_{1\le
    i <j\le 4}(-1)^{i+j+1}x_i^V\Theta^{-1}x_j^V}
\eeqa
\noi
where $x_1^V,\ldots, x_4^V$ are the $4-$vectors of the positions of the $4$
fields incident to the respective vertex $V$.
%To any such vertex one associates 
%a hypermomentum $p_V$ using
%the relation
%\beqa
%\label{pbar1}
%\delta(x_1^V -x_2^V+x_3^V-x_4^V ) 
%%&=&
%= \int  \frac{d p_V}{(2 \pi)^4}
%e^{p_V \sigma (x_1^V-x_2^V+x_3^V-x_4^V)}.
%\eeqa

\medskip

To end this section, let us remark that the Moyal space, as a linear space of infinite dimension, admits a
particular base, the matrix base, which in two dimensions can be indexed by
two natural numbers. All the NCQFT entities expressed in this section can be
rewritten in this base (see for example \cite{GW1, GW2}). However, we do not
introduce it here, since this is not requested to present our results. 

\section{Parametric representation}
%\label{defnc}
%\resetequ
\setcounter{equation}{0}

%when considering the parametric
%representation for 
In the case of 
commutative QFT, one has translation invariance in position
space. As a consequence of this invariance, the first polynomial vanishes when
integrating over all internal positions. Therefore, one has to integrate over
all internal positions (which correspond to vertices) save one, which is thus
singularized. However, the polynomial is a still a canonical object, {\it
  i. e.} it does not depend of the choice of this
particular vertex.

%As noticed in \cite{param1}, i
In the non-commutative case the translation
invariance is lost (see previous section). Therefore, one can integrate
over all internal positions and hypermomenta, without vanishing of the first
polynomial. However, in order to be able to recover the commutative limit, we
also singularize a particular vertex. We call this particular vertex the {
  root}. Because there is no translation invariance, the polynomial
does  depend on the choice of the root; however the leading UV 
terms do not.

From the propagator \eqref{propa} and the vertices
contributions \eqref{v1} one is able to write the Feynman amplitude ${\cal A}$ as function of
the non-commutative polynomials $HU$ and $HV$ as
\beqa
\label{HUGV}
{\cal A} = K \int_{0}^{1} \prod_{\ell=1}^L  [ d t_\ell
(1-t_\ell^2) ]
HU ( t )^{-\frac D2}   
e^{-  \frac {HV (t)}{HU ( t )}},
\eeqa
where  $K$ is some constant, unessential
for this calculus and by $x_e$ and $D$ we mean the external positions of the
graph and resp. the space-time dimension. 
In \cite{param1} it was furthermore proved that  $HU$ and $HV$
are polynomials in the set of variables $t$.

Let us state that, even the formulas
above hold also for non-orientable graphs (that is graphs corresponding to
interactions $\bar \phi \star \bar \phi \star \phi \star \phi$), for simplicity reasons 
we restrict ourselves
to the study of polynomials for orientable graphs (that is graphs corresponding to
interactions $\bar \phi \star \phi \star \bar \phi \star \phi$, as already
mentioned in the previous section).

\subsection{The parametric representation for the Grosse-Wulkenhaar model}

In \cite{param1}, non-zero {\it leading terms} ({\it i. e.} terms which have
the smallest global degree in the $t$ variables) of $HU$ were identified. 
These terms are dominant in the UV regime.

In order to characterize some of them, we need the following definition:

\begin{definition}
\label{admisibila}
Let  a subset $J$ of the set $\{1, \ldots, L\}$ of internal lines of a
Feynman graph. Then $J$ is a hyper-tree if
%\begin{itemize}
%\item 
it contains a tree  in the dual graph
 and  its complement contents a tree  in the direct Feynman graph.
%\end{itemize}
\end{definition}
Let  $|J|$ be the cardinal of the set $J$.
Considering now a Feynman graph of genus $g$ and $F$ faces.
%one is able to 
%set a lower limit on the polynomial $HU$:
In \cite{param1} it was proven the theorem:

\begin{theorem}
One has the following lower limit on the polynomial $HU$
\beqa
\label{limit}
HU (t)  \ge 
\sum_{J\, hyper-tree} (2s)^{2g-k_{J}} \prod_{\ell\in J} t_\ell,
\eeqa
where $s=\frac{1}{\Omega}$ and $k_J=|J|-F+1$.
\end{theorem}

\subsection{Parametric representation for the Langmann-Szabo-Zarembo Model}
%\label{defnc}

%Let now $K$
%be a subset of $\{1,\ldots,L\}$. Let
%\beqa
%\label{kij2}
%k_{K} = \vert K\vert- L - F +1
%\eeqa
%and $n_{K}=\mathrm{Pf}(B'_{\hat{K}})$, the Pfaffian of the  matrix
%$B'$  
%with deleted lines and columns $K$ among the first $L$ indices 
%(corresponding to short variables $u$).

%The specific form \eqref{defmatrixa2}
%allows, as in the previous subsection, to write the polynomial $HU$ as a sum of positive terms:
%\beqa
%\label{suma2}
%HU_{G,{\bar V}} (t) &=&  \sum_{K}  s^{2g-k_{K}} \ n_{K}^2
%\prod_{\ell \in K} \frac{1+t_\ell^2}{2t_\ell} \prod_{\ell' \in \{1,\ldots, L\}} t_{\ell'}\ .
%\eeqa

%As for the Grosse-Wulkenhaar model, i
It was proven in \cite{param2} that one
can compute some leading terms for this type of model also. 
%Indeed, 
%when
%choosing a subset of internal lines 
%$ K= \{1, \ldots, L\} - J_0, $
%where $J_0$ is an admissible set (as defined in the previous subsection), one has
%$$ n_K= 2^g\prod 2 (\Omega \pm 1)$$
%(the product of the factors $(\Omega \pm 1)$ depending on the topology of the
%graph). 
%As in the previous subsection 

\begin{theorem}
One has the following lower limit on the polynomial $HU$
\beqa
\label{limit2}
HU (t) \ge  \sum_{J_0\, {\rm hyper-tree}}  s^{2[g+(F-1)]} \ 
\left( 2^g \prod 2 (\Omega \pm 1) \right)^2\nonumber\\
\prod_{\ell \in K} \frac{1+t_\ell^2}{2t_\ell} \prod_{\ell' \in \{1,\ldots,L\}} t_{\ell'},
\eeqa
where $K= \{1, \ldots, L\} - J_0,$, with $J_0$ some admissible set.
\end{theorem}
%and this is the main result of \cite{param2}. 

Note that the product of the factors $(\Omega \pm 1)$ depends on the
topology of the graph (see \cite{param2} for details). 

\medskip

Theorem \ref{limit} or Theorem \ref{limit2} allow to obtain the
following power counting for both these models
\beqa
\label{power}
 \omega = 4g + \frac 12 (N-4),
\eeqa
where $\omega$ is the superficial degree of convergence and $N$ is the number of external legs of the respective graph.

Let us now make some comments on the results of this section. First of all,
one notices an improvement in the power counting \eqref{power}, improvement
given by the presence of a new term in the graph genus.  Moreover, let us recall that, in commutative QFT 
the parametric representation leads naturally to the topological notion of
trees and to some ``democracy'' between them (one sums over all trees, with
the same weight for each of them). The
non-commutative 
equivalent of these properties is the natural appearance of the more involved
topological notion of hyper-trees and a corresponding ``democracy'' between them.
Another important issue to stress on is (as in the commutative case) the
explicit positivity of the formulas. Finally, let us state that in all
the formulas of the non-commutative parametric representation the space-time
dimension $D$ is just a parameter. It is again the exact same situation as for the
parametric representation for commutative QFT.

\medskip

Note that for both type of models, when considering second polynomial $HV$,  similar leading $UV$ terms, similar results
of positivity, boundness, 
``democracy'' between adapted topological entities and finally  
power counting 
have been obtained. 
%in
%\cite{param1} and resp. \cite{param2}  for the second polynomial $HV$ too.

\section{Mellin representation; meromorphy in $D$}
%\label{defnc}
%\resetequ
\setcounter{equation}{0}
 
Following \cite{mellin} we present here the Mellin representation for the
Feynman amplitudes of a graph corresponding to the Grosse-Wulkenhaar or the
LSZ model. 
The polynomial $HU$ can be written as
\beqa
\label{mu}
HU=\sum_{{K_U}}a_{K_U} \prod_{\ell=1}^L t_\ell^{u_{\ell {K_U}}}=\sum_{{K_U}}HU_{{K_U}},
\eeqa
where $K_U$ is a reunion of subsets of internal lines, $a_{K_U}$ is some
constant (depending on the topology) and $u_{\ell {K_U}}$ is
an exponent which can take the values $0,1$ or $2$ (see \cite{mellin} for
details). The difference with the commutative case comes from the presence of
the constants  $a_{K_U}$ as well as from the fact that the exponents $u_{\ell
  {K_U}}$ are allowed to take the value $2$.
 
The second polynomial $HV$ has both a real $HV^\ccR$ and an imaginary part
$HV^\cI$. This also is a major difference with respect to the commutative
case. 
One now writes down formulas  analogous to \eqref{mu}, formulas involving 
the
reunion of subsets of internal lines $K_V$, 
the monomials $HV_{K_V}^\ccR$ and $HV_{K_V}^\cI$, 
the constants $s_{K_V}^\ccR$ and
$s_{K_V}^\cI$ and the exponents $v_{\ell K_V}$.

In order to introduce the Mellin representation, one writes for the real part
$HV^\ccR$ of $HV$ 
%one uses again the identity \eqref{gamayk} as
\begin{equation} 
\label{yR}
e^{-HV^\ccR_{K_V}/U}=\int_{\tau_{K_V}^\ccR}\Gamma (-y_{K_V}^\ccR)\left( \frac{HV^\ccR_{K_V}}U\right) ^{y_{K_V}^\ccR}, 
\end{equation} 
where $\int_{\tau_{K_V}^\ccR}$
is a short notation for 
$\int_{-\infty }^{+\infty }% 
\frac{d(\cI\, y_{K_V})}{2\pi }$, 
with $\ccR\, y_{K_V}$ fixed at $\tau _{K_V}^\ccR<0$.
This formula introduces the set of Mellin parameters $y_{K_V}^\ccR$.

A similar formula is written for the imaginary part $HV^\cI$ of $HV$, which
introduces  the set of Mellin parameters $y_{K_V}^\cI$. Note that in this case
this will hold in the sense of distributions. This comes from the fact that
the non-commutative vertex contribution (see \eqref{v1}) has a distributional
form. This is the major difference with respect to the commutative case.

For the  polynomial $HU$ one has
\begin{equation} 
\Gamma \left( \sum_{K_V}y_{K_V}^\ccR+y_{K_V}^\cI+\frac D2\right) (HU)^{-\sum_{K_V} (y_{K_V}^\ccR+ y_{K_V}^\cI) -\frac D2}=\int_\sigma 
\prod_{K_U}\Gamma (-x_{K_U})U_{K_U}^{x_{K_U}},  \label{x} 
\end{equation}
where $\int_\sigma$ is a short notation for 
 $\int_{-\infty }^{+\infty }\prod_{K_U}
\frac{d(\cI\, x_{K_U})}{2\pi }$ with $\sum_{K_U}x_{K_U}+\sum_{K_V} (y_{K_V}^\ccR+y_{K_V}^\cI) =-\frac D2$.
Furthermore, let 
\begin{equation} 
\phi_\ell = \sum_{K_U} u_{\ell {K_U}}x_{K_U}+\sum_{K_V}
(v^\ccR_{\ell {K_V}}y_{K_V}^\ccR+v^\cI_{\ell {K_V}}y_{K_V}^\cI) +1 .
\end{equation} 
and the convex domain
\begin{equation} 
\Delta =\left\{ \sigma ,\tau^\ccR, \tau^\cI \left|  
\begin{array}{l} 
\sigma _{K_U}<0;\;\tau^\ccR_{K_V}<0;\; -1 < \tau^\cI_{K_V}<0;\;
\\
\sum_{K_U}x_{K_U}+\sum_{K_V} (y_{K_V}^\ccR+ y_{K_V}^\cI)=-\frac D2; \\  
\forall \ell,\;\ccR  \, \phi_\ell\equiv \sum_{K_U}u_{\ell {K_U}}\sigma _{K_U}\\
+\sum_{K_V}
(v^\ccR_{\ell {K_V}}\tau^\ccR_{K_V}+v^\cI_{\ell {K_V}}\tau^\cI_{K_V})+1>0 
\end{array} 
\right. \right\}  \label{nc-domain} 
\end{equation} 
where $\sigma $, $\tau^\ccR $ and $\tau^\cI $ stand 
for $\ccR \, x_{K_U}$, $\ccR  \, y_{K_V}^\ccR$   and $\ccR  \, y_{K_V}^\cI$.

Putting all these together, one is able to prove (see again \cite{mellin}) the
following theorem:

\begin{theorem}\label{cmrep}
A Feynman amplitude of a Grosse-Wulkenhaar or LSZ graph
is analytic in the strip $0<\Re \, D<2$ where it writes
\beqa 
\label{final}
{\cal A}_\cG =
%&=&
{\rm K'}\int_\Delta \frac{\prod_{K_U} a_{K_U}^{x_{K_U}} \Gamma (-x_{K_U})}{\Gamma (-\sum_{K_U}x_{K_U})}
\left( \prod_{K_V} (s_{K_V}^\ccR)^{y_{K_V}^\ccR}\Gamma (-y_{K_V}^\ccR) \right)
%\nonumber\\
%&& 
\left( \prod_{K_V} (s_{K_V}^\cI)^{y_{K_V}^\cI}\Gamma (-y_{K_V}^\cI) \right)
\left( \prod_{\ell=1}^L \frac {\Gamma (\frac{\phi_\ell}2)
  \Gamma (\frac D2)}{2\Gamma (\frac{\phi_\ell+D}2)} \right).
\eeqa
where $\int_\Delta $ is a short notation for  integration over the
variables $\frac{\cI\, x_{K_U}}{2\pi i}$, $\frac{\cI\, y_{K_V}^\ccR}{2\pi i}$
 and $\frac{\cI\, y_{K_V}^\cI}{2\pi i}$ in the domain $\Delta$. 
\end{theorem}

%Note that t
This theorem holds as tempered distribution of the external
invariants. It is this the main difference with the commutative case:  this integral representation (previously true 
in the sense of \textit{functions} of the external invariants)
now holds only in the sense of {\it distributions}.  Indeed,
the distributional character of commutative amplitudes 
reduces to a single overall $\delta$-function of momentum conservation. 
This is no longer true
for these non-commutative amplitudes, which must be seen as distributions smeared
against test functions of the external variables.

Furthermore, let us note here that the representation given by Theorem
\ref{cmrep} allows the study of asymptotic behavior under rescaling of
arbitrary subsets of external invariants of a Feynman amplitude.

\medskip

Finally, let us end this paper by a theorem regarding the meromorphy of a
Feynman amplitude in $D$:

\begin{theorem}\label{meromorph}
Any Feynman amplitude is a tempered meromorphic distribution in $D$,
{\it i. e.} 
 the amplitude smeared against any fixed Schwarz-class test function
of the external invariants yields a meromorphic function in $D$
in the entire complex plane, with singularities located among a discrete rational 
set which depends only on the graph and not on the test function. 
%In the $\phi^4$ case, no such singularity can occur in the strip $0<\Re D<2$.  
%a region which is therefore a germ of analyticity common to the whole theory.
\end{theorem}

The fact that all the formulas in this paper present $D$ as a simple
parameter, as well as Theorem \ref{meromorph} above, pave the road for dimensional
regularization and renormalization of these theories.

\end{document}